\documentclass{iopart}

\usepackage{url}
\usepackage{amsbsy}
\usepackage{graphicx}
\usepackage[usenames]{color}
\usepackage{iopams}
\usepackage{multirow}
\usepackage{hyperref}

\newcommand{\eqref}[1]{{(\ref{#1})}}
 
\begin{document}

\title{The Mock LISA Data Challenges: from Challenge 3 to Challenge 4} 

\author{The \emph{Mock LISA Data Challenge Task Force}:
Stanislav Babak$^1$,
John G. Baker$^2$,
Matthew J. Benacquista$^3$,
Neil J. Cornish$^4$,
Shane L. Larson$^5$,
Ilya Mandel$^6$,
Sean T. McWilliams$^2$,
Antoine Petiteau$^1$,
Edward K. Porter$^7$,
Emma L. Robinson$^1$,
Michele Vallisneri$^{8,9}$,
Alberto Vecchio$^{10}$,
and \emph{the Challenge 3 participants}:
Matt Adams$^4$,
Keith A. Arnaud$^2$,
Arkadiusz B{\l}aut$^{11}$,
Michael Bridges$^{12}$,
Michael Cohen$^9$,
Curt Cutler$^{8,9}$,
Farhan Feroz$^{12}$,
Jonathan R. Gair$^{13}$,
Philip Graff$^{12}$,
Mike Hobson$^{12}$,
Joey Shapiro Key$^4$,
Andrzej Kr\'olak$^{14}$,
Anthony Lasenby$^{12,15}$,
Reinhard Prix$^{16}$,
Yu Shang$^1$,
Miquel Trias$^{17}$,
John Veitch$^{10}$,
John T. Whelan$^{18}$}
\address{$^1$ Max-Planck-Institut f\"ur Gravitationsphysik (Albert-Einstein-Institut), Am M\"uhlenberg 1, D-14476 Golm bei Potsdam, Germany}
\address{$^2$ Gravitational Astrophysics Lab., NASA Goddard Space Flight Center, 8800 Greenbelt Rd., Greenbelt, MD 20771, USA}
\address{$^3$ Center for Gravitational Wave Astronomy, Univ.\ of Texas at Brownsville, Brownsville, TX 78520, USA}
\address{$^4$ Dept.\ of Physics, Montana State Univ., Bozeman, MT 59717, USA}
\address{$^5$ Dept.\ of Physics, Utah State Univ., Logan, UT 84322, USA}
\address{$^6$ Dept.\ of Physics and Astronomy, Northwestern Univ., Evanston, IL, USA}
\address{$^7$ APC, UMR 7164, Univ.\ Paris 7 Denis Diderot, 10, rue Alice Domon et Leonie Duquet, 75025 Paris Cedex 13, France}
\address{$^8$ Jet Propulsion Laboratory, California Inst.\ of Technology, Pasadena, CA 91109, USA}
\address{$^9$ Theoretical Astrophysics, California Inst.\ of Technology, Pasadena, CA 91125}
\address{$^{10}$ School of Physics and Astronomy, Univ.\ of Birmingham, Edgbaston, Birmingham B152TT, UK}
\address{$^{11}$ Inst.\ of Theoretical Physics, Univ.\ of Wroc{\l}aw, Wroc{\l}aw, Poland}
\address{$^{12}$ Astrophysics Group, Cavendish Laboratory, Univ.\ of Cambridge, Cambridge CB30HE, UK} 
\address{$^{13}$ Inst.\ of Astronomy, Univ.\ of Cambridge, Cambridge CB30HA, UK}
\address{$^{14}$ Inst.\ of Mathematics, Polish Academy of Sciences, Warsaw, Poland}
\address{$^{15}$ Kavli Institute for Cosmology, Univ.\ of Cambridge, Cambridge CB30HA, UK}
\address{$^{16}$ Max-Planck-Institut f\"{u}r Gravitationsphysik (Albert-Einstein-Institut), D-30167 Hannover, Germany}
\address{$^{17}$ Departament de F\'isica, Univ.\ de les Illes Balears, Cra.\ Valldemossa Km.\ 7.5, E-07122 Palma de Mallorca, Spain}
\address{$^{18}$ Center for Computational Relativity and Gravitation \& School of Math.\ Sciences, Rochester Inst.\ of Technology, 85 Lomb Memorial Drive, Rochester, NY 14623, USA}

\ead{Michele.Vallisneri@jpl.nasa.gov}

\begin{abstract}
The Mock LISA Data Challenges are a program to demonstrate LISA data-analysis capabilities and to encourage their development. Each round of challenges consists of one or more datasets containing simulated instrument noise and gravitational waves from sources of undisclosed parameters. Participants analyze the datasets and report best-fit solutions for the source parameters. Here we present the results of the third challenge, issued in Apr 2008, which demonstrated the positive recovery of signals from chirping Galactic binaries, from spinning supermassive--black-hole binaries (with optimal SNRs between $\sim 10$ and $2000$), from simultaneous extreme--mass-ratio inspirals (SNRs of 10--50), from cosmic-string--cusp bursts (SNRs of 10--100), and from a relatively loud isotropic background with $\Omega_\mathrm{gw}(f) \sim 10^{-11}$, slightly below the LISA instrument noise.
\end{abstract}

\vspace{-18pt}


\section{The Mock LISA Data Challenges}

The Laser Interferometer Space Antenna (LISA) is a planned NASA--ESA gravitational-wave (GW) observatory sensitive in the $10^{-5}$--$10^{-1}$ Hz range \cite{lisa}. LISA's data will contain superposed signals from millions of sources, including all the binaries in the Galaxies with orbital periods below five hours and massive--black-hole (MBH) binary coalescences out to $z \sim 20$ \cite{baker2007}. Thousands of sources will be resolvable individually. The potential for source confusion and the very complex dynamics and waveforms of sources such as extreme--mass-ratio inspirals (EMRIs), suggested the need for a coordinated effort to develop and demonstrate LISA data-analysis capabilities. The Mock LISA Data Challenges (MLDCs) began in early 2006 with this very purpose.

The complexity and ambition of the challenges has risen with each round: MLDC 1 \cite{mldclisasymp,mldcgwdaw1} focused on simple sources, isolated or moderately interfering; MLDC 2 \cite{mldcgwdaw2,mldcamaldi2} introduced a Galactic \emph{ensemble} of 26 million binaries (20,000 of which were successfully recovered), as well as the problem of detecting MBH binaries over the Galactic cacophony; MLDC 1B \cite{MLDC3} reprised the first challenge for new research collaborations joining the effort, and saw the first successful detections of EMRI signals. MLDC 3, released in April 2008 and due in April 2009, consisted of five subchallenges that featured more realistic models of previously examined sources (chirping Galactic binaries in MLDC 3.1, spinning MBH binary inspirals in 3.2, superposed EMRIs in 3.3) and entirely new sources (GW bursts from cosmic-string cusps in 3.4, an isotropic stochastic background in 3.5); see \cite{MLDC3} for a detailed discussion of the source models and GW content of each subchallenge. Fifteen collaborations, comprising all the participants listed in the byline and most task-force members, submitted a total of seventeen entries (all can be found at \url{www.tapir.caltech.edu/~mldc/results3} together with technical notes about search implementation).

In this paper we briefly report on the detection and parameter-estimation performance demonstrated by each entry. Altogether, MLDC 3 showed substantial progress in tackling increasingly difficult data-analysis problems, and introduced new search methods such as nested sampling and sophisticated genetic optimization--Markov Chain hybrids. However, there is certainly room for improvement and further work: fewer Galactic binaries were recovered by the searches employed here than by the multi-source MCMC demonstrated in MLDC 2; MBH binaries and EMRIs were detected with high confidence, but the accurate estimation of their parameters (beyond the dominant ones) was stymied by the complex global structure and the many local maxima of likelihood surfaces. On the bright side, searches for cosmic-string bursts and stochastic backgrounds (admittedly simple problems in the absence of nonstationary or non-Gaussian instrument noise) met no roadblocks.

Section \ref{sec:mldc4} of this article introduces MLDC 4, which is being released as we write (November 2009), with entries due at the end of 2010.

\section{Galactic Binaries (MLDC 3.1)}

Challenge dataset 3.1 contained signals from over 60 million
chirping Galactic binaries. The vast majority of these are too
weak to be isolated, and the unresolved component forms a
nonstationary confusion noise that adds to the overall noise level.
Estimates based on self-consistent removal schemes \cite{Timpano:2005gm}
and Bayesian model selection \cite{Crowder:2007ft} suggest that it should be
possible to recover between 20,000 and 30,000 binaries. Three groups submitted source catalogs for MLDC 3.1:
\begin{itemize}
\item \textbf{BhamUIB} (a collaboration between the Universities of
Birmingham and the Balearic Islands) implemented a delayed-rejection MCMC algorithm \cite{trias2009} to search three narrow frequency windows, $0.3 \, {\rm mHz} \leq f \leq 0.4 \, {\rm mHz}$,
$0.9 \, {\rm mHz} \leq f \leq 1.0 \, {\rm mHz}$, and
$1.6 \, {\rm mHz} \leq f \leq 1.7 \, {\rm mHz}$, using the MLDC waveform
generator \cite{Cornish:2007if}. A total of 494 sources were reported.
\item \textbf{AEIRIT} (researchers at the Albert Einstein Institute in Hannover, Germany, and the Rochester Institute of Technology) set up a LIGO-style
hierarchical search based on the ${\cal F}$-statistic and on frequency-domain rigid-adiabatic templates \cite{whelan2009}.
Triggers are generated for the individual TDI channels; those found in coincidence
are analyzed coherently using noise-orthogonal TDI observables. A total of 1940 sources
were reported.
\item \textbf{PoWrWa} (a collaboration between the Albert Einstein Institute, the University of Wroc{\l}aw, and the Polish Academy of
Sciences) adopted an iterative matched-filtering search that used the ${\cal F}$-statistic and rigid-adiabatic templates \cite{babaknew}, and analyzed a few 0.1-mHz wide frequency bands. The brightest source in each band is identified and
removed, and the process repeated until a pre-set SNR threshold is
reached. A total of 14,838 sources were reported.
\end{itemize}
The entries
for this round fell short of the theoretical target for a variety of reasons.
BhamUIB analyzed only a small fraction of the data,
while AEIRIT and PoWrWa used single-pass or iterative search
schemes, which are limited in how deep they can dig before source confusion
degrades signal recovery. Previous studies~\cite{Crowder:2007ft}
indicate that it should be possible to recover approximately 99\% of
the resolvable sources to an accuracy of better than 90\%, as measured
by the overlap between injected and recovered waveforms. We therefore
adopt correlation as the metric by which the entries are measured.

For this we need to identify the injected signal that corresponds most closely to each recovered signal. To do this, we consider all injected signals with ${\rm SNR} > 3$ within six frequency bins of the recovered signal, and we select the injected signal with the minimum $\chi^2 = (h-h' \vert h-h')$,
where $h$ and $h'$ are the injected and recovered waveforms, and $(a\vert b)$ denotes the standard noise-weighted inner product, summed over the noise-orthogonal $A$, $E$, and $T$ TDI channels. The correlation is then computed as $C = (h \vert h')/
(h \vert h)^{1/2} (h' \vert h')^{1/2}$. Using as a figure of merit the percentage of recovered sources with $C > 0.9$, we find 30\% for BhamUIB, 95\% for AEIRIT, 33\% for PoWrWa. PoWrWa have reported a bug in their code that affects sources with frequencies above 3 mHz. Results are better after correcting it \cite{babaknew}. If we consider only the 6,955 sources they report below 3 mHz, their figure of merit improves to 58\%.

\section{Massive black-hole binaries (MLDC 3.2)}

Challenge dataset 3.2 contained five signals from MBH inspirals, embedded in instrument noise and a partially subtracted Galactic background (see \cite{MLDC3} for a synopsis of the waveform model and for the random selection of source parameters). Multi-TDI SNRs ranged from $\sim 13$ to $\sim 1671$, with the two weakest signals (MBH-2 and MBH-6) corresponding to mergers after the end of the dataset. Five groups submitted entries:
\begin{itemize}
\item \textbf{AEI} (researchers at the Albert Einstein Institute) used a genetic matched-filtering algorithm extended to a multi-modal search; as a merit function they took the $A$-statistic, a geometrical mean of the log-likelihood for the signal and for a low-frequency signal subset \cite{GAspinbbhFullPaper}.
\item \textbf{CambAEI} (a collaboration between the Cambridge Cavendish Laboratory and Institute of Astronomy with the the Albert Einstein Institute) adopted the MultiNest implementation of nested-sampling integration \cite{MultiNest2} to compute the evidence and produce posterior distributions. This method is intrinsically multimodal, and the $A$-statistic was used to identify modes.
\item \textbf{GSFC} (researchers at Goddard Space Flight Center) employed the tempered--Metropolis-Hastings MCMC algorithm found in Xspec \cite{Xspec}. 
\item \textbf{JPLCITNWU} (a collaboration between JPL/Caltech and Northwestern University) employed a two-stage method whereby they followed up a nonspinning MBH search \cite{JPLCaltech} with MultiNest.
\item \textbf{MTGWAGAPC} (researchers at Montana State University and at the Astro-Particle and Cosmology Institute, APC, in Paris) adopted a parallel-tempering MCMC algorithm, evolved from the thermostated, frequency-annealed algorithm employed in previous challenges \cite{SMBHCornishPorter}.
\end{itemize}
All five groups recovered parameters for the loud signals; MBH-2 (SNR $\sim$ 19) was recovered by AEI, CambAEI, JPLCITNWU and MTGWAGAPC; MBH-6 (SNR $\sim$ 13) by AEI, CambAEI and MTGWAGAPC. Participants were encouraged to submit multiple modes of comparable probability, if present. For the best (highest-SNR) mode in each entry, Table \ref{tab:SMBH_Err} lists the fractional parameter errors, as well as the recovered SNR and individual TDI-channel fitting factors (FFs). Both the recovered SNR and the $\mathrm{SNR}_\mathrm{true}$ of the true waveform were computed by filtering the noisy dataset (in its LISA Simulator version) with the appropriate template, while the FFs were computed between noiseless signals. Note that the reported modes had often SNR differences of less than one, and the highest-SNR mode did not always have the best parameters.
\begin{table}
\caption{Selected parameter errors, SNRs, and FFs for each group's highest-SNR entries to MLDC 3.2. The time of coalescence $t_c$, spin magnitudes $a_{1,2}$ and luminosity distance $D$ are defined in Table 7 of \cite{MLDC3}; in addition, the (redshifted) chirp mass $M_c \equiv (m_1 m_2)^{3/5} / (m_1 + m_2)^{1/5}$, and the symmetric mass ratio $\eta = m_1 m_2 / (m_1 + m_2)^{2}$. $\Delta\textrm{sky}$ is the angular geodesic distance between 
the estimated and true positions; values $\sim 180$ deg correspond to the antipodal sky location, a known quasi-degeneracy in the LISA response.\label{tab:SMBH_Err}} \vspace{-12pt}
\lineup \scriptsize \flushright
\begin{tabular}{l@{\;}r@{\;}|@{\;}r@{\;}r@{\;}r@{\;}r@{\;}r@{\;}r@{\;}r@{\;}|@{\;}r@{\;}r@{\;}r}
\br
source & group & $\Delta M_{c}/ M_{c}  $& $\Delta \eta/ \eta $ & $ \Delta t_{c} $ &  $ \Delta\mathrm{sky}$ & $ \Delta a_{1} $ & $ \Delta a_{2}  $ &  $\Delta D / D$ & SNR & $\mathrm{FF}_A$ & $\mathrm{FF}_E$ \\
$(\mathrm{SNR}_\mathrm{true})$ & & $\times 10^{-5}$ & $\times 10^{-4}$ & (sec) & (deg) & $\times 10^{-3}$ & $\times 10^{-3}$ & $\times10^{-2}$   \\
\mr
\textbf{MBH-1}              & AEI            &         2.4 &        6.1 &   62.9 &   11.6 &    7.6 &   47.4 &   8.0 & 1657.71 & 0.9936 & 0.9914 \\
(1670.58)              & CambAEI &         3.4 &     40.7 &   24.8 &      2.0 &    8.5 &   79.6 &   0.7 & 1657.19 & 0.9925  & 0.9917   \\
         & MTAPC    &       24.8 &     41.2 & 619.2 & 171.0 & 13.3 &   28.7 &    4.0  & 1669.97 & 0.9996 & 0.9997 \\
              & JPL           &       40.5 &  186.6 &   23.0 &    26.9 & 39.4 &   66.1 &    6.9  & 1664.87 & 0.9972 & 0.9981 \\
              & GSFC       &  1904.0 &  593.2 & 183.9 &    82.5 &   5.7 & 124.3 &  94.9  &  267.04 & 0.1827 & 0.1426 \\
\mr
\textbf{MBH-3}              & AEI            &       9.0 &         5.2 &    100.8 & 175.9 &      6.2 &    18.6 &   2.7 & 846.96 & 0.9995 & 0.9989  \\
(847.61)              & CambAEI &     13.5 &      57.4 &    138.9 & 179.0 &    21.3 &      7.2 &   1.5 & 847.04  & 0.9993 & 0.9993 \\
         & MTAPC    &   333.0 &    234.1 &    615.7 &   80.2 &    71.6 & 177.2 & 16.1 & 842.96  & 0.9943 & 0.9945  \\
              & JPL           &   153.0 &      51.4 &    356.8 &   11.2 & 187.7 & 414.9 &    2.7 & 835.73 & 0.9826 & 0.9898  \\
              & GSFC       & 8168.4 & 2489.9 & 3276.9 &    77.9 & 316.3 &   69.9 &  95.6 & 218.05 & 0.2815 & 0.2314 \\

\mr
\textbf{MBH-4}              & AEI            &      4.5 &      75.2 &       31.4 &   0.1 & 47.1 &173.6  &    9.1 & 160.05 & 0.9989 &  0.9994 \\
(160.05)              & CambAEI &      3.2 &    171.9 &      30.7 &   0.2 & 52.9 & 346.1 &  21.6  & 160.02 & 0.9991 &  0.9992  \\
         & MTAPC    &     48.6 & 2861.0 &        5.8 &   7.3 & 33.1 & 321.1 & 33.0  & 149.98  & 0.8766 &  0.9352  \\
              & JPL           &  302.6 &    262.0 &   289.3 &   4.0 & 47.6 & 184.5 & 28.3  & 158.34  & 0.8895 &  0.9925 \\
              & GSFC       &  831.3 & 1589.2 & 1597.6 & 94.4 & 59.8 & 566.7 & 95.4 & $-45.53$ & $-0.1725$ & $-0.2937$ \\

\br
\textbf{MBH-2}              & AEI            & 1114.1 & 952.2 & 38160.8 & 171.1 & 331.7 & 409.0 &  15.3 & 20.54 & 0.9399 & 0.9469 \\
(18.95)         & CambAEI &      88.7 & 386.6 &   6139.7 & 172.4 & 210.8 & 130.7 &  24.4  & 20.36 & 0.9592 & 0.9697 \\
              & MTAPC    &   128.6 &   45.8 & 16612.0 &      8.9 & 321.4 & 242.4 &  13.1  & 20.27 & 0.9228 & 0.9260 \\
              & JPL           &   287.0 & 597.7 & 11015.7 &   11.8 & 375.3 & 146.3 &    9.9 & 18.69 & 0.9661 & 0.9709 \\
\mr
\textbf{MBH-6}             & AEI            &    1042.3 & 1235.6 &   82343.2 &      2.1 & 258.2 & 191.6 & 26.0 & 13.69 &  0.9288 &  0.9293 \\
(12.82)         & CambAEI &    5253.2 & 1598.8 & 953108.0 & 158.3 & 350.8 & 215.4 & 29.4 & 10.17 &  0.4018 &  0.4399 \\
              & MTAPC    & 56608.7 &    296.7 & 180458.8 & 119.7 & 369.2 & 297.6 & 25.1 & 11.34 & -0.0004 &  0.0016 \\
\br
\end{tabular}
\end{table}

\begin{figure}
\includegraphics[width=\textwidth]{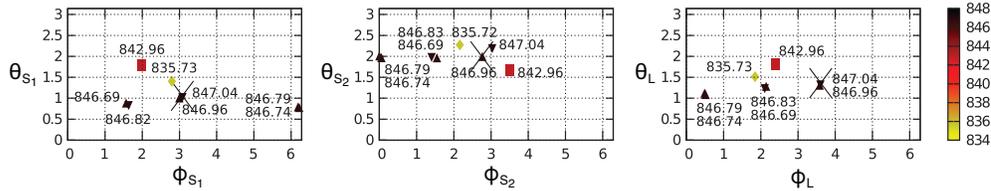} \\[-18pt]
\caption{Distribution of reported modes for MBH-3 along the spin and orbital--angular-momentum angles. Each mode is annotated with the SNR.
$\times$: true value; $\blacktriangle$: AEI; $\blacktriangledown$: CambAEI; $\blacksquare$: MTGWAGAPC; $\blacklozenge$: JPLCITNWU.\label{fig:SMBH_spinLdeg}} \vspace{-12pt}
\end{figure}

The binaries that merge within the duration of the dataset (MBH-1, 3, and 4) have significantly larger SNR and reach higher frequencies, and thus allow a much more accurate recovery of source parameters. For these binaries, the mass, time-of-coalescence, distance, and sky-position errors are comparable, if not better, to those of nonspinning-binary searches (see Tables 1 and 2 of \cite{mldcamaldi2}). The errors in the spin amplitudes are consistent with the Fisher-matrix results of \cite{SpinBBHLangHughes}.
However, despite the high SNR, the estimation of the initial direction of the spins and of the orbital angular momentum has proved to be very difficult. We observe a large number of the local likelihood maxima that have comparable, high SNR (and $\mathrm{FF} > 0.99$), yet very different values of these parameters, as illustrated in Fig.\ \ref{fig:SMBH_spinLdeg} for the MBH-3 entries.

Parameter estimation was not as successful for the weaker signals with mergers beyond the end of the dataset. Still, for MBH-2 the errors in the masses and time of coalescence are comparable to the Fisher-matrix predictions. The errors in sky position are $\sim 10$ deg, with strong local likelihood maxima at the antipodal sky position. Spin amplitudes are determined very poorly; this reflects the fact that the spins are nearly degenerate with other parameters in the low-frequency part of the waveforms.

Lang and Hughes \cite{SpinBBHLangHughes} report that spin-induced modulations remove correlations between parameters in Fisher-matrix computations, improving overall parameter determination. However, here we observe that spin interactions also cause nonlocal degeneracies in parameters space, especially so for spin and orbital--angular-momentum angles. Further investigations are needed to determine which phenomenon is stronger. Nevertheless, the entries to MLDC 3.2 demonstrate a solid detection capability for spinning-binary inspirals, and a good recovery of most source parameters.

\section{Extreme--mass-ratio inspirals (MLDC 3.3)}


Challenge dataset 3.3 contained five Barack--Cutler \cite{barackcutler} EMRI signals immersed in instrument noise (see \cite{MLDC3} for details about the waveforms and the random choice of parameters). In comparison to previous EMRI challenges, here participants had to contend with multiple simultaneous sources, as well as weaker signals---the injected SNRs varied between $20$ and $37$. Three groups submitted entries:
\begin{itemize}
\item \textbf{BabakGair} (a collaboration between the Albert Einstein Institute and Cambridge University) used stochastic sampling and MCMC to identify EMRI harmonics, then carried out an $\mathcal{F}$-statistic search in the space of harmonics, and performed a final MCMC fit in source parameter space. This search improved on the method described in \cite{BabakGairPorter}, adding more sophisticated harmonic identification.
\item \textbf{EtfAG} (researchers at Cambridge and Northwestern Universities) searched for harmonics in the time--frequency spectrogram using the Chirp-based Algorithm for Track Search (CATS, \cite{CATS}) developed for earlier MLDCs \cite{GairMandelWen}, and improved for this search to deal with intersecting tracks from multiple sources.
\item \textbf{MTAPCIOA} (a collaboration between Montana State University, APC--Paris, and Cambridge University) improved the MLDC algorithm used in previous rounds \cite{Cornish:2008} to include parallel tempering as described in \cite{keycornish}, and to enhance the implementation of ``harmonic jumps'' between secondary likelihood maxima. 
\end{itemize}
Parameter-estimation errors are presented in Table \ref{tab:EMRI_Err}.
Altogether, all EMRIs were found by at least one group, and masses were estimated accurately; more work remains to be done on improving the estimates of EMRI parameters for relatively weak and overlapping signals.

MTAPCIOA recovered all five EMRI signals, generally with very good parameter-estimation accuracies: errors of a few tenths \% in the masses of both bodies, and sky-position errors of a few deg ($13$ for EMRI-4). Their second solution for EMRI-1 was particularly impressive, with fractional errors of a few $10^{-5}$ in masses, initial eccentricity, and spin. However, their other solution exemplifies the difficulty of resolving secondary maxima: the SNR is almost the same ($21.794$ vs.\ $21.804$), but parameter errors are two orders of magnitude greater.
\begin{table}
\caption{Parameter-estimation errors for the EMRIs in MLDC 3.3. $M$ and $\mu$ are the masses of the central and inspiraling bodies; $\nu_0$ and $e$ are the initial azimuthal orbital frequency and eccentricity; $|S|$ is the dimensionless central-body spin; $\lambda_\mathrm{SL}$ is the spin--orbit misalignment angle, and $D$ the luminosity distance. $\Delta$spin and $\Delta$sky are the geodesic angular distances between the estimated and true spin direction and sky position. $\mathrm{SNR}_\mathrm{true}$ is computed with the LISA Simulator; the SNR for each entry with the simulator used in that search (the LISA Simulator \cite{lisasimulator} for MTAPCIOA, Synthetic LISA \cite{synthlisa} for EtfAG and BabakGair).\label{tab:EMRI_Err}} \vspace{-12pt}
\lineup \scriptsize \flushright
\begin{tabular}{l@{\;}l|r@{\;}r@{\;}r@{\;}r@{\;}r@{\;}r@{\;}r@{\;}r@{\;}r@{\;}r}
\br
Source & Group & SNR & $\frac{\Delta M}{M}$ & $\frac{\Delta \mu}{\mu}$ & $ \frac{\Delta \nu_0}{\nu_0} $ &  $\Delta  e_0$ &  $\Delta |S|$ & $\frac{\Delta \lambda_{\rm SL}}{\lambda_{\rm SL}}$ & $\Delta$spin & $\Delta$sky & $\frac{\Delta D}{D}$ \\
($\mathrm{SNR}_\mathrm{true}$) & & & $\times 10^{-3}$ & $\times 10^{-3}$ & $\times 10^{-5}$ & $\times 10^{-3}$  & $\times 10^{-3}$ & $\times 10^{-3}$ & (deg) & (deg) &    \\
\mr
\textbf{EMRI-1} & MTAPCIOA & 21.794 & 5.05  &  3.29 &  1.61  &  $-5.1$\0  &  $-1.4$\0  &  $-19$\,\0\0  &  23\,\0  &  2.0  &  0.07  \\ 
(21.673) & MTAPCIOA & 21.804 & $-0.06$  &  $-0.01$  &  $-0.08$  &  $-0.05$  &  0.02  &  0.54  &  3.5  &  1.0  &  0.13 \\ 
\mr
\textbf{EMRI-2} & MTAPCIOA & 32.387 & $-3.64$  &  $-2.61$  &  $-3.09$  &  3.8\0  &  0.87  &  12\,\0\0  &  11\,\0  &  3.7  &  $3\!\times\!10^{-3}$  \\ 
(32.935) & BabakGair & 22.790 & 33.1\0  &  $-19.7$\0  &  10.1\0  &  $-33$\,\0\0  &  $-7.3$\0  &  250\,\0\0  &  47\,\0  &  3.5  &  $-0.25$  \\  
& BabakGair & 22.850 & 32.7\0  &  $-20.0$\0  &  9.94  &  $-32$\,\0\0 &  $-7.2$\0  &  250\,\0\0  &  58\,\0  &  3.5  &  $-0.24$  \\ 
& BabakGair & 22.801 & 33.5\0  &  $-19.5$\0  &  10.5\0  &  $-33$\,\0\0  &  $-7.4$\0  &  240\,\0\0  &  40\,\0  &  3.5  &  $-0.25$  \\     
\mr
\textbf{EMRI-3} & MTAPCIOA & 19.598 & 1.62  &  0.38 &  $-0.10$  &  $-0.35$  &  $-0.94$  &  $-3.0$  &  5.0  &  3.0  &  $-0.04$  \\  
(19.507) & BabakGair & 21.392 & 1.77  &  1.01  &  1.95  &  $-1.2$\0 &  $-0.68$  &  $-2.3$  &  116\,\0  &  4.5  &  0.13  \\   
& BabakGair & 21.364 & 2.26  &  1.88  &  2.71  &  $-2.0$\0  &  $-0.69$  &  $-2.5$  &  65\,\0  &  6.1  &  0.14  \\    
& BabakGair & 21.362 & 1.51  &  1.01  &  2.09  &  $-1.3$\0  &  $-0.50$  &  $-1.7$  &  7.6  &  6.2  &  0.14  \\  
& EtfAG & --- & 54.0\0 &  4.88  &  $-7375$\,\0\0  &  26\,\0\0  &  17\,\0\0  &  ---  &  ---  &  32\,\0  &  0.83  \\        
\mr
\textbf{EMRI-4} & MTAPCIOA & $-0.441$ & $-8.77$  &  $-10.1$\0  &  $-6.03$  &  $-3.7$\0 &  144\,\0\0  &  950\,\0\0  &  99\,\0  &  13\,\0  &  $-2.3$\0  \\ 
(26.650) & \\
\mr
\textbf{EMRI-5} & MTAPCIOA  & 17.480  &  $-3.32$  &  5.00  &  $-1.80$  &  0.22  &  55\,\0\0  &  62\,\0\0  &  43\,\0  &  1.8  &  $-1.3$\0  \\ 
(36.173) & \\
\br                                              
\end{tabular}
\vspace{-12pt}
\end{table}

The time--frequency analysis carried out by EtfAG was particularly hard-hit by the simultaneous lowering of the SNR and the presence of multiple overlapping signals in MLDC 3.3.  The group was unable to find the low-frequency, low-SNR EMRI-1; while EtfAG did find a medium-frequency source, the relatively large parameter-estimation errors suggest that the time--frequency approach did not adequately resolve between the overlapping harmonics of EMRI-2 and EMRI-3.

BabakGair submitted three (relatively close) solutions for each of EMRI-2 and EMRI-3.  For EMRI-3, their estimates are better than those of EtfAG and comparable to those of MTAPCIOA, albeit somewhat less accurate for the initial parameter values and distance.  For EMRI-2, BabakGair had errors of a few \% in the masses and initial eccentricity, and had significant errors in spin-orientation angle and in distance, although sky location was still found correctly, within $3.5$ deg.

The high-frequency EMRI-4 and EMRI-5 presented a challenge for all groups.  Although MTAPCIOA found them and estimated their masses fairly accurately, the errors in spin magnitude and orientation were significantly larger than for other sources, and the distance to both sources was overestimated by factors of $2$ or $3$ (vs.\ errors $\lesssim 10\%$ for the other EMRIs).  Furthermore, the negative SNR for claimed EMRI-4 and the low FFs between the recovered and injected noiseless waveforms
indicate that the MTAPCIOA search could not resolve these sources individually, but converged on two parameter sets that jointly fit the combination of the two injected sources.   

\section{Cosmic-string--cusp bursts (MLDC 3.4)}
\begin{table}
\caption{Select parameter errors, SNRs, and FFs for the MLDC 3.4 entries. Here $\Delta$sky is the geodesic angular distance between the recovered and true sky positions; $t_D$ is the time of burst arrival at LISA; $\psi$ and $\mathcal{A}$ are the GW polarization and amplitude. CaNoE reported only $t_C$ and $\mathcal{A}$, so $\Delta$sky, $\Delta t_D$, $\Delta \psi$, SNR, and FFs cannot be computed.\label{tab:parerrs}}
\lineup \scriptsize \flushright
\begin{tabular}{l@{\;\;}r@{\;\;}|@{\;\;}r@{\;\;}r@{\;\;}r@{\;\;}r@{\;\;}|@{\;\;}r@{\;\;}r@{\;\;}r}
\br
source & group & $\Delta$sky & $\Delta t_D $ & $ \Delta \psi $ &  $ \Delta{\mathcal A}/{\mathcal A}$  & SNR & $\mathrm{FF}_A$ & $\mathrm{FF}_E$ \\
$(\mathrm{SNR}_\mathrm{true})$ & & (deg) & (sec) & (rad) &  &  &  &    \\
\mr
\textbf{String-1}              & CAM            &         106.9 &        1.462 &   0.501 &   0.904  & 43.706 & 0.99947 & 0.99797 \\
(43.46)              & CAM &         49.4 &     2.331 &   1.065 &      1.128  & 43.520 & 0.99964  & 0.99591   \\
              & JPLCIT           &       34.2 &  1.585 &   3.726 &    0.413   & 43.506 & 0.99986 & 0.99844 \\
              & JPLCIT       &  113.7 &  1.574 & 3.739 &    0.431   &  43.497 & 0.99988 & 0.99847 \\
              & MTGWAG           &       106.6 &  2.071 &   2.600 &    0.745   & 43.287 & 0.99975 & 0.99565 \\
\mr
\textbf{String-2}              & CAM            &       82.0 &         3.683 &    4.846 & 0.062  & 33.690 & 0.99945 & 0.99986  \\
(33.6)         & JPLCIT    &   90.5 &    4.005 &    4.268 &   0.282  & 33.689  & 0.99949 & 0.99929  \\
              & JPLCIT           &   45.2 &      3.847 &    6.364 &   0.231  & 33.694 & 0.99939 & 0.99960  \\
              & MTGWAG       & 53.1 & 3.223 & 0.158 &    0.011  & 33.696 & 0.99926 & 0.99978 \\

\mr
\textbf{String-3}              & CAM            &      80.8 &      1.249 &       3.785 &  0.338  & 41.326 & 0.99073 &  0.99923 \\
(41.42)              & CAM &      133.3 &    1.715 &      3.257 &   0.238   & 41.456 & 0.99388 &  0.99869  \\
         & CAM    &     44.5 & 0.763 &        3.202 &   0.066   & 41.142  & 0.99700 &  0.99883  \\
              & JPLCIT       &  59.0 & 1.546 & 3.129 & 0.317  & 41.315 & 0.99554 & 0.99848 \\
              & JPLCIT           &  157.7 &    1.226 &   5.614 &   0.220   & 41.316  & 0.99717 &  0.99864 \\
              & MTGWAG       &  137.9 & 0.980 & 0.110 & 0.161  & 41.418 & 0.99327 & 0.99948 \\
\br
\end{tabular}
\vspace{-12pt}
\end{table}

Challenge dataset 3.4 contained three burst signals from cosmic-string cusps, immersed in instrument noise with slightly randomized levels for each individual noise (i.e., from the six proof masses and photodetectors). The dataset was less than a month long ($2^{21}$ s), with a higher sampling rate (1 s) than the others, to accommodate the potential high-frequency content of these signals, which have power-law spectrum up to an $f_\mathrm{max}$ determined by the characteristic length scale of the string and the viewing angle (see \cite{MLDC3} for more details about the waveforms and the random choice of their parameters). Four collaborations submitted entries:
\begin{itemize}
\item \textbf{CAM} (a collaboration between Cambridge U.\ and APC--Paris) used MultiNest.
\item \textbf{CaNoe} (researchers at Cambridge and Northwestern Universities) implemented a time--frequency algorithm, a modified version of CATS \cite{CATS}.
\item \textbf{JPLCIT} (Caltech/JPL) experimented with MCMC and MultiNest, but only submitted entries based on the latter \cite{cohen2009}.
\item \textbf{MTGWAG} (Montana State University) used a parallel-tempering MCMC \cite{keycornish}. 
\end{itemize}
All groups successfully recovered all three bursts. Table \ref{tab:parerrs} shows the parameter-estimation errors, SNRs, and FFs for all entries. 
Although the accuracy of parameters is poor, the FFs are very high; this suggests that these results are not due to shortcomings in the search methods, but to the very character of the waveforms. As a matter of fact, for relatively short signals such as these bursts, LISA can be considered as a static detector, and its response is not imprinted with any modulations from the LISA orbit or from the rotation of its constellation. Hence the sky position of burst sources can only be determined by triangulation between the spacecraft---a weaker effect, and one that vanishes in the limit of long wavelengths.

Thus, the determination of sky position is intrinsically harder for bursts, and it is further complicated by the presence of degeneracies \cite{keycornish} such as the reflection of sky position across the instantaneous LISA plane. The bursts' barycentric central time $t_C$ and its polarization are strongly coupled with sky position, and therefore are also determined poorly. To compensate for this fact, we calculate the error in the arrival time of the burst at the center of the detector constellation $t_D$. This parameter has a weaker correlation with the sky position and constrained better by observations.

\section{Stochastic background (MLDC 3.5)}

\begin{figure}
\lineup \scriptsize \flushright
\begin{minipage}{0.5\textwidth}
\vspace{-12pt}
\includegraphics[width=\textwidth]{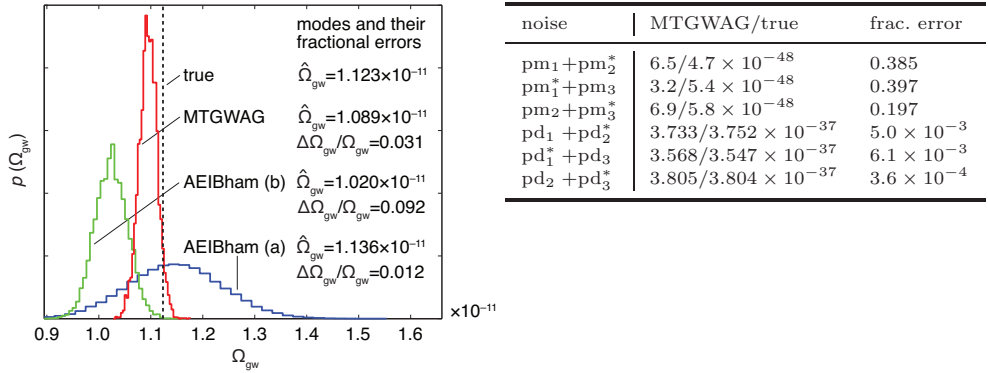}
\end{minipage}
\begin{tabular}[b]{l@{+}l|ll}
\br
\multicolumn{2}{l|}{noise} & MTGWAG/true & frac.\ error \\
\mr
pm$_1$ & pm$_2^*$ & $6.5/4.7 \times 10^{-48}$ & $0.385$ \\ 
pm$_1^*$ & pm$_3$ & $3.2/5.4 \times 10^{-48}$ & $0.397$ \\ 
pm$_2$ & pm$_3^*$ & $6.9/5.8 \times 10^{-48}$ & $0.197$ \\ 
pd$_1$ & pd$_2^*$ & $3.733/3.752 \times 10^{-37}$ &  $5.0 \times 10^{-3}$ \\ 
pd$_1^*$ & pd$_3$ & $3.568/3.547 \times 10^{-37}$ &  $6.1 \times 10^{-3}$ \\ 
pd$_2$ & pd$_3^*$ & $3.805/3.804 \times 10^{-37}$ &  $3.6 \times 10^{-4}$ \\ 
\br
\end{tabular}
\vspace{-12pt}
\caption{Plot on left: posterior PDFs for $\Omega_\mathrm{gw}$ reported by each analysis, their modes $\hat{\Omega}_\mathrm{gw}$, and the corresponding fractional error w.r.t.\ the true $\Omega_{\mathrm{gw}}$. The horizontal-axis range corresponds to the ``prior'' MLDC range. Table on right: estimated LISA noise levels (modes) in the MTGWAG analysis. The $\mathrm{pm}_k$ and $\mathrm{pm}^*_k$ are the proof mass noises introduced in the left and right optical assemblies on LISA spacecraft $k$; likewise for photodetector noises $\mathrm{pd}_k$ and $\mathrm{pd}^*_k$. The combinations of noises shown in the table are the only ones that are constrained effectively by the data after laser phase noise has been removed with TDI.\label{fig:stochastic_pdf}}
\vspace{-12pt}
\end{figure}

Challenge dataset 3.5 contained an isotropic stochastic background signal immersed in instrument noise, with noise levels slightly randomized as in MLDC 3.4.
To facilitate the use of the GW-canceling TDI combination $T$, 
the LISA orbits were approximated as those traced by a rigidly rotating
triangle, with equal and constant arm lengths (except for the Sagnac effect). The background was realized by placing $192 \times 2$ linearly polarized, stochastic pseudosources at positions uniformly distributed across the sky. See \cite{MLDC3} for more details. Now, isotropic stochastic backgrounds can be characterized by a single dimensionless quantity,
\begin{equation}
\Omega_{\mathrm{gw}}(f)=\frac{1}{\rho_{\mathrm{crit}}}\frac{\mathrm{d} \rho_{\mathrm{gw}}(f)}{\mathrm{d} \log{f}},
\end{equation}
where $\rho_{\mathrm{gw}}(f)$ is the energy density in GWs,
and $\rho_{\mathrm{crit}}=3c^2H_0^2/8\pi G$ is the energy density required
to close the Universe. In this MLDC round, $\Omega_{\mathrm{gw}}$ was taken to be constant across frequencies, and in this dataset it was set equal to $1.123 \times 10^{-11}$ (with $H_0=70 \mathrm{km}/\mathrm{s}/\mathrm{Mpc}$). Two groups submitted entries for this challenge, and both analyzed the same version of the dataset, generated by Synthetic LISA \cite{synthlisa}:
\begin{itemize}
\item \textbf{AEIBham} (a collaboration between the Albert Einstein Institute
and the University of Birmingham) submitted multiple entries using different applications of the same MCMC algorithm \cite{Robinson2008}. Here we focus on the AEIBham analysis of the TDI $A$ and $E$ observables, which was run separately over frequency bands of 0.1--1 mHz (a) and 0.1--5 mHz (b).
\item \textbf{MTGWAG} (Montana State University) analyzed the TDI $A$, $E$, and $T$ spectra with parallel-tempering MCMC. They estimated the background level as well as the levels of select linear combinations of the individual LISA noises (other combinations are left essentially undetermined by TDI observations).
\end{itemize}
Figure \ref{fig:stochastic_pdf} shows the reported posterior PDFs and the best-fit values for $\Omega_{\mathrm{gw}}$, along with their fractional error; both groups recovered the injected value within less than $10\%$. Figure \ref{fig:stochastic_pdf} shows also the best-fit levels reported by MTGWAG for the determinable proof-mass and photodetector noise combinations.
The AEIBham analyses used theoretical expressions for TDI $A$ and $E$ noise spectra, which left only their overall levels as unknowns. However, this treatment assumed that the secondary noises were symmetric, which was not the case in this dataset. In addition, because of limitations in the MLDC stochastic-background generation code, the actual GW strain spectrum differed from the nominal $f^{-3}$ power law at higher frequencies.
This may explain why the AEIBham (b) result was further from the injected
value that either the AEIBham (a) result (in which band the GW spectrum was actually $f^{-3}$) or the MTGWAG result, which used all three TDI channel, solved for their unknown correlations, and calibrated the GW spectrum using training datasets.

\section{Moving forward: a synopsis of MLDC 4}
\label{sec:mldc4}

While the third round of the MLDCs was focused on increasing the complexity and variety of GW sources, we are devoting the next iteration to the \emph{global-fit} problem of detecting and analyzing sources of different types superposed in the LISA data. Thus, MDLC 4 consists of a single, ``whole enchilada'' challenge that includes all the sources of MLDCs 3.1--3.5 \emph{in the same dataset}, albeit with larger source numbers (for EMRIs and cosmic-string bursts) and parameter ranges (for MBH binaries and EMRIs). See Table \ref{table:MLDC4} for details. The duration of the dataset is again approximately two years ($2^{22} \times 15$ s), but the data are released both as a high-cadence time series of $2^{25}$ samples with $\Delta t = 1.875$ s, and as a downsampled time series of $2^{22}$ samples with $\Delta t = 15$ s (however, only the cosmic-string bursts and instrument noise contribute spectral content above 33 mHz).

Three versions of both datasets are released: fractional-frequency--fluctuation data generated by Synthetic LISA \cite{synthlisa} and LISACode \cite{lisacode}, and equivalent-strain data generated by the LISA Simulator \cite{lisasimulator}. In contrast to MLDCs 3.4 and 3.5, all the LISA instrument noises are set to symmetric, nominal levels, and the LISA armlength are allowed to ``breathe.'' See \cite{mldcgwdaw2,MLDC3} for more details about the MLDC models of LISA orbits, noises, and interferometric observables. The waveform models of MLDC 3 \cite{MLDC3} are used unchanged in MLDC 4, except for the following: the spectra of cosmic-string bursts are truncated below $10^{-5}$ Hz; the stochastic-background spectrum is $f^{-3}$ between $10^{-5}$ Hz and 33 mHz, and drops lower below and above that range; last, 
when MBH binaries end within the duration of the dataset, they are terminated 
at the time $t_\mathrm{max}$ when the frequency derivative changes sign, indicating that the PN expansion is failing. To reduce spectral leakage, a half Hann window $\cos^2[\pi (t - t_\mathrm{max} + \Delta t) / 2 \Delta t]$ is applied between $t_\mathrm{max} - \Delta t$ and $t_\mathrm{max}$, with $\Delta t = Q/f_\mathrm{max}$, $f_\mathrm{max}$ the frequency at time $t_\mathrm{max}$, and $Q$ the quality factor (lisaXML parameter \texttt{Q}) set to 3 for MLDC 4 (setting \texttt{Q}=0 yields the old $r = 6M$ termination condition).
\begin{table}
\caption{MLDC 4 source content and parameter priors (cf.\ Tables 7 and 8 of \cite{MLDC3}; the parameters in \textbf{bold} have changed compared to MLDC 3). Parameters are always sampled uniformly from the ranges given below; angular parameters are drawn from appropriate uniform distributions on the circle or sphere. Source distances are set from the SNRs selected randomly in the ranges given below, defining $\mathrm{SNR} = \sqrt{2} \times \mathrm{max} \{\textrm{SNR}_X,\textrm{SNR}_Y,\textrm{SNR}_Z\}$. The notation $\mathbf{Poisson}(\lambda)$ indicates a random number of sources distributed according to a Poisson distribution with mean $\lambda$.\label{table:MLDC4}}
\lineup \scriptsize \flushright
\begin{tabular}{l@{\hspace{6pt}}l}
\br
\textit{Galactic-binary background} & $\sim 34 \times 10^6$ interacting, $\sim 26 \times 10^6$ detached systems \\
\mr
4--6 \textit{MBH binaries} & $m_1 = \mathbf{0.5}\mbox{--}5 \times 10^6\,M_\odot$, $m_1/m_2 = 1\mbox{--}\mathbf{10}$, $a_1/m_1 = 0\mbox{--}1$, \\
& $a_2/m_2 = 0\mbox{--}1$, with $t_c$ and SNRs as in MLDC 3.2 \\
\mr
an average of 6 \textit{EMRIs} & $\mu = 9.5\mbox{--}10.5 \, M_\odot$, $S = 0.5\mbox{--}0.7 \, M^2$, $e_\mathrm{plunge} = \mathbf{0.05}\mbox{--}0.25$, \\
                                             & $t_\mathrm{plunge} = 2^{21}\mbox{--}2^{22} \times 15$ s, SNR $= \mathbf{25\mbox{--}50}$ \\
\multicolumn{1}{r}{\ldots including}         & \textbf{Poisson(2)} systems with $M = 0.95\mbox{--}1.05 \times 10^7 M_\odot$ \\
& \textbf{Poisson(2)} systems with $M = 4.75\mbox{--}5.25 \times 10^6 M_\odot$ \\
& \textbf{Poisson(2)} systems with $M = 0.95\mbox{--}1.05 \times 10^6 M_\odot$ \\
\mr
\textbf{Poisson(20)} \textit{cosmic-string bursts} & $f_\mathrm{max} = 10^{-3\mbox{--}1} \, \mathrm{Hz}$, $t_C = \mathbf{0\mbox{--}2^{22} \times 15}$ s, $\textrm{SNR} = 10\mbox{--}100$ \\
\mr
\textit{isotropic stochastic background} & $S^\mathrm{tot}_h = 0.7\mbox{--}1.3 \times 10^{-47} (f/\mathrm{Hz})^{-3} \, \mathrm{Hz}^{-1}$ \\
\br
\end{tabular}
\vspace{-12pt}
\end{table}

Challenges beyond MLDC 4 will feature ever more realistic datasets, including more sophisticated waveform models, such as MBH coalescences with merger and ringdown phases, and ``real-mission'' noise models and instrument events drawn from the ongoing experimental investigations of the LISA architecture and subsystem. To obtain more information and to participate in the MLDCs, see the official MLDC website, \url{astrogravs.nasa.gov/docs/mldc}, the Task Force wiki, \url{www.tapir.caltech.edu/listwg1b}, and the \texttt{lisatools} software repository \cite{lisatools}.

\ack The authors acknowledge support from
the German Aerospace Center (SB, AP, JTW);
the Max-Planck Society (SB, JTW, AP, RP, ELR);
the LISA Mission Science Office at JPL (CC, MV);
MNiSW grant N N203 387237 (AB, AK);
NASA grants NNG04GD52G (MB),
  NNG05GI69G and NNX07AJ61G (NJC),
  06-BEFS06-31 (MV and CC),
  NNX07AH22G (IM),
  and the NASA Center for GW Astronomy at the U. of Texas, Brownsville;
NSF grant PHY-0855494 (JTW),
  and the NSF A\&A postdoctoral fellowship under AST-0901985 (IM);
the Science and Technology Facilities Council (AV);
St Catharine's College, Cambridge (JG);
E.U.\ FEDER funds, Spanish Ministerio de Ciencia e Innovaci\'on projects FPA2007-60220, HA2007-0042 and CSD2007-00042, Govern de les Illes Balears, Conselleria d'Economia, Hisenda i Innovaci\'o (MT).
CC's and MV's work was carried out at the Jet Propulsion Laboratory, California Institute of Technology, under contract with the National Aeronautics and Space Administration. Copyright 2009. All rights reserved.


\section*{References}

\begin{thebibliography}{99}
%
\bibitem{lisa} Bender P, Danzmann P and the LISA Study Team 1998 ``Laser Interferometer Space Antenna for the Detection of Gravitational Waves, Pre-Phase A Report'' \textbf{MPQ 233} (Garching: Max-Planck-Instit\"ut f\"ur Quantenoptik) 
%
\bibitem{baker2007} Baker J G et al. 2007 \textit{Phys. Rev. D} \textbf{75} 124024
%
\bibitem{mldclisasymp} Arnaud K A et al. (the MLDC Task Force) 2006 \textit{Laser Interferometer Space Antenna: 6th International LISA Symp. (Greenbelt, MD, 19--23 Jun 2006)} ed Merkowitz S M and Livas J C (Melville, NY: AIP) p 619; \textit{ibid} p 625 (long version with lisaXML description: gr-qc/0609106)
%
\bibitem{mldcgwdaw1} Arnaud K A et al. (the MLDC Task Force and Challenge 1 participants) 2007 \textit{Class. Quant. Grav.} \textbf{24} S529
%
\bibitem{mldcgwdaw2} Arnaud K A et al. (the MLDC Task Force) 2007 \textit{Class. Quant. Grav.} \textbf{24} S551
%
\bibitem{mldcamaldi2} Babak S et al. (the MLDC Task Force and Challenge 2 Participants) 2008 \textit{Class. Quant. Grav.} \textbf{25} 114037
%
\bibitem{MLDC3} Babak et al. 2008 \textit{Class. Quant. Grav.} \textbf{25} 184026
%
\bibitem{Timpano:2005gm} Timpano S E, Rubbo L J and Cornish N J 2006 \textit{Phys. Rev. D} \textbf{73} 122001
%
\bibitem{Crowder:2007ft} Crowder J and Cornish N J 2007 \textit{Class. Quant. Grav.}  \textbf{24} S575
%
\bibitem{trias2009} Trias M, Vecchio A and Veitch J 2009 arXiv:0904.2207 
%
\bibitem{Cornish:2007if} Cornish N J and Littenberg T B 2007 \textit{Phys. Rev. D} \textbf{76} 083006
%
\bibitem{whelan2009} Whelan J T, Prix R and Khurana D 2009 arXiv:0908.3766;
2008 \textit{Class. Quantum Grav.} \textbf{25} 184029;
Prix R and Whelan J T 2007 \textit{Class. Quant. Grav.} \textbf{24} S565 
%
\bibitem{babaknew} B{\l}aut A, Babak S and Kr{\'o}lak A 2009 arXiv:0911.3020
%
\bibitem{GAspinbbhFullPaper} Petiteau A, Shang Y, Babak B and Feroz F \textit{in preparation}
%
\bibitem{MultiNest2} Bridges M, Feroz F and Hobson M P 2009 \textit{MNRAS} \textbf{398} 1601
%
\bibitem{Xspec} Arnaud K A et al., Xspec website, \url{http://heasarc.gsfc.nasa.gov/xanadu/xspec}
%
\bibitem{JPLCaltech} Brown D A et al. 2007 \textit{Class. Quant. Grav.} \textbf{24} S595  
%
\bibitem{SMBHCornishPorter} Cornish N J and Porter E K 2006 \textit{Class. Quant. Grav.} \textbf{24} 5729
%
\bibitem{SpinBBHLangHughes} Lang R N and Hughes S A 2006 Phys. Rev. D \textbf{74} 122001
%
\bibitem{barackcutler} Barack L and Cutler C 2004 \textit{Phys. Rev.} D \textbf{69} 082005
%
\bibitem{BabakGairPorter} Babak S, Gair J R and Porter E 2009 \textit{Class. Quant. Grav.} \textbf{26} 135004.
%
\bibitem{CATS} Mandel I and Gair J R \textit{in preparation}
%
\bibitem{GairMandelWen} Gair J R, Mandel I and Wen L 2008 \textit{Class. Quant. Grav.} \textbf{25} 184031
%
\bibitem{Cornish:2008} Cornish N J 2008 arXiv:0804.3323
%
\bibitem{keycornish} Shapiro Key J and Cornish N J 2009 \textit{Phys. Rev. D} \textbf{79} 043014
%
\bibitem{lisasimulator} Cornish N J and Rubbo L J 2003 \textit{Phys. Rev. D} \textbf{67} 022001; erratum-ibid. 029905; LISA Simulator website, \url{www.physics.montana.edu/lisa}; included in \cite{lisatools}
%
\bibitem{synthlisa} Vallisneri M 2005 \textit{Phys. Rev. D} \textbf{71}; Synthetic LISA website, \url{www.vallis.org/syntheticlisa}; included in \cite{lisatools}
%
\bibitem{cohen2009} Cohen M, Cutler C and Vallisneri M \textit{in preparation}
%
\bibitem{Robinson2008} Robinson E L, Romano J D and Vecchio A 2008 \textit{Class. Quant. Grav.} \textbf{25} 184019 
%
\bibitem{lisacode} Petiteau A et al. 2008 \textit{Phys. Rev. D} \textbf{77} 023002; LISACode website, \url{www.apc.univ-paris7.fr/LISA-France/analyse.phtml}; included in \cite{lisatools}
%
\bibitem{lisatools} LISAtools website and SVN repository, \url{lisatools.googlecode.com}

\end{thebibliography}
\end{document}